%Paper: hep-th/9310050
%From: PARANJ@LPS.UMONTREAL.CA
%Date: Thu, 7 Oct 1993 16:18:54 -0400 (EDT)

\magnification 1200
\def\entier{{\rm Z}\mskip-6mu{\rm Z}}

\def\reel{{\rm I}\mskip-3.5mu{\rm R}}
\def\complexe{\lower-0.33ex\hbox{${\scriptstyle |}$}\mskip-8mu{\rm C}}

\vsize=22truecm
\hsize=15truecm
\voffset=0.5truecm
\hoffset=1truecm

\pageno=1
\baselineskip=18truept
\lineskip=2truept
\lineskiplimit=0truept
\line{\hfil hep-th/yymmxxx}
\line{\hfil UdeM-LPN-TH-93-176}
\vskip 0.5cm
\centerline{22$^{th}$ Conference on Differential Geometrical Methods in
Theoretical Physics}
\vskip 0.5cm
\centerline{Ixtapa, Mexico, September,1993}
\vskip 2.0cm
\centerline{\bf SKYRMION-SKYRMION SCATTERING}
\vskip 0.5cm
\centerline{by T. Gisiger and M. B. Paranjape}
\vskip 0.5cm
\centerline{\it Laboratoire de physique nucl\'eaire, Universit\'e de
Montr\'eal}
\vskip 0.5cm
\centerline{\it C.P. 6128, succ ``A", Montr\'eal, Qu\'ebec, Canada, H3C 3J7 }
\vskip 0.8cm
\centerline{\bf Abstract}
\vskip 1.0cm
We study the scattering of Skyrmions at low energy and large separation using
the method proposed by Manton of truncation to a finite number of degree
freedom. We calculate the induced metric on the manifold of the union of
gradient flow curves, which for large separation, to first non-trivial order is
parametrised by the variables of the product ansatz.

\vfill\break

\vskip 1.0cm
\noindent {\bf 1. Introduction}
\vskip 1.0cm

Scattering of solitons in a non-integrable, non-linear classical or quantum
field theory remains an intractable and difficult problem, however, it
concerns one
of the most interesting
aspects of the nature of the corresponding physics. Numerical methods have
given reasonable ideas on how the scattering proceeds but they are still
unsatisfactory for uncovering the detailed dynamics governing the scattering.

A method has been proposed by Manton$^1$ for truncating the degrees of freedom
from the original infinite number to a relevant finite number of variables. The
idea first considers the case of theories of the Bogomolnyi type, those
theories which admit {\it static} soliton solutions, usually in the topological
two soliton sector, which
asymptotically describe two single solitons at arbitrary positions and relative
orientations. The configuration at small separation contains, in general,
strong
deformations of the individual solitons and in fact they lose their identity.
However the set of configurations have the same energy since they correspond
to the continuous variation of a
finite number of parameters, the modulii.  Otherwise they could not be
stationary points of the potential.
In general, for solitons corresponding to a topological quantum number, the
modulii space corresponds to the sub-manifold of minimum energy configurations
within the given topological sector. Manton suggests that the low energy
scattering of solitons, with initial configuration on this sub-manifold
corresponding to asymptotic, single solitons, with arbitrarily small initial
velocity tangent to the sub-manifold, will self-consistently be constrained
to remain on the sub-manifold. Since the potential energy is a constant on the
sub-manifold the resulting dynamics reduces to geodesic motion on the
sub-manifold in the induced metric on the sub-manifold from the
kinetic term.
It is a difficult task to prove such a truncation of degrees of freedom
in a mathematically rigorous fashion, however, it does seem intuitively
correct. The non-linearity of the theory implies the coupling of the degrees of
freedom corresponding to the sub-manifold with all other excitations through
the potential. We are assuming that these are negligible. Manton and
Gibbons$^2$ applied this program with remarkable success to the case of
magnetic monopoles in the BPS limit and it has also been applied to vortex
scattering in a similar limit$^3$.

The generalization to the more common situation where the set of static
solutions correspond to a finite set of critical points proceeds as follows.
The critical points are typically a minimum energy configuration which is
essentially a bound state of two solitons, an asymptotic critical point which
corresponds to two infinitely separated solitons and possibly a number of
unstable non-minimal critical points of varying energies of the same order.
These critical points are degenerate with a finite number of degrees of
freedom. They are connected by special paths, the paths of steepest
descent or equivalently the
gradient flow curves. In this case Manton proposes that the dynamics will be
constrained to lie on the sub-manifold comprising of the union of all these
curves. This again is intuitively reasonable. If we think of the space of all
configurations as a large bag, the bottom surface of the bag will correspond to
this sub-manifold, and a slow moving marble rolling on the bottom will tend to
stay there.

The Skyrme model falls into the second case. We identify the corresponding
sub-manifold for well-separated Skyrmions and we calculate the induced metric
to lowest non-trivial inverse order in the separation from the kinetic term.
This is the first step towards calculating the scattering of Skyrmions
in this formalism.

\vskip 2.0cm
\noindent {\bf 2. The Skyrme model}
\vskip 1.0cm

The Skyrme model is described by the Lagrangean,
$${\cal L} =
{f_\pi^2 \over 4} tr(U^\dagger \partial_\mu U U^\dagger \partial^\mu U)
+ {1\over 32 e^2}\,tr( [U^\dagger \partial_\mu U,U^\dagger \partial_\nu U]^2)$$
where $U(x)$ is a unitary matrix valued field. We take
$$U(x) \in SU(2).$$
The Skyrme Lagrangean corresponds to first terms of a systematic expansion in
derivatives of the effective Lagrangean describing describing low energy
interaction of pions. It is derivable from QCD hence $f_\pi$ and $g$ are in
principle calculable from QCD. What is even more surprising is that it includes
the baryons as well which arise as topological solitonic solutions of the
equations of motion. The original proposal of this by Skyrme$^4$ in the 60's
was put on solid footing by Witten$^5$ in the 80's.

The topological solitons, called Skyrmions, correspond to non-trivial mappings
of $\reel^3$ plus the point at infinity into $SU(2)$:
$$U(x): \reel^3 + \infty \to SU(2) = S^3.$$
But
$$\reel^3 + \infty = S^3$$
thus the homotopy classes of mappings
$$U(x): S^3 \to S^3$$
which define
$$\Pi_3(S^3) = \entier$$
characterize the space of configurations.

The topological charge of each sector is given by
$$ N = {1 \over 24 \pi^2} \int{d^3 \vec x\, \epsilon^{ijk}
tr(U^\dagger \partial_i U U^\dagger \partial_j U U^\dagger \partial_k U)}$$
which is identified with the baryon number. Thus for the scattering of two
Skyrmions, we are looking at the sector of baryon number equal to 2.
In this sector the minimum energy configuration should correspond to the bound
state of two Skyrmions, which must represent the deuteron. The asymptotic
critical point corresponds to two infinitely separated Skyrmions.
There exist, known, non-minimal critical points, corresponding to a spherically
symmetric configuration, the di-baryon solution$^6$. The energy of this
configuration is about three times the energy of a single Skyrmion. There
are also, possibly, other non-minimal critical points with energy less than two
infinitely separated Skyrmions$^7$. The scattering of two
Skyrmions will take place
on the union of the paths of steepest descent which connect the various
critical points.

\vskip 2.0cm
\break
\noindent {\bf 3. Skyrmion-Skyrmion Scattering}
\vskip 1.0cm

We consider the scattering only for large separation. In this way we do not
have to know the structure of this manifold in the complicated
region where the two Skyrmions interact strongly and consequently are much
deformed. In the region of large separation the product ansatz corresponds to
$$
\eqalign{
U(\vec x) &= U_1(\vec x - \vec R_1) U_2(\vec x - \vec R_2)\cr
          &= A_1^\dagger U(\vec x - \vec R_1)A_1A_2^\dagger
 U(\vec x - \vec R_2)A_2}
$$
where $U(\vec x - \vec R_1)$ and $U(\vec x - \vec R_2)$ correspond to the field
of a single Skyrmion solution centered at
$R_1$ and $R_2$ respectively. The full Skyrme model dynamics implies a
deformation of each Skyrmion. This deformation, from a numerical studies, is
found to be unimportant already at a separation of 1.5 fermi$^8$. We will
neglect this deformation.

It remains to calculate the metric on the sub-manifold parametrized by the
product ansatz. We find the interesting result that the metric behaves like
$1/d$ where $d$ is the separation$^9$. We find the kinetic energy:
$$T = - 2 M + {1\over 2} M \dot{\vec R_1}^2 + {1\over 2} M \dot{\vec R_2}^2
- \Lambda \,tr(A^{\dagger}_1 \dot A_1 A^{\dagger}_1 \dot A_1)
- \Lambda \,tr(A^{\dagger}_2 \dot A_2 A^{\dagger}_2 \dot A_2) + T_{int}$$
where
$$M = 4\pi \int_0^\infty r^2dr{ \biggl\{ {1\over 8} f_\pi^2
\biggl[\biggl({\partial f\over \partial
r}\biggr)^2\!\!+ 2\,{\sin^2 f\over r^2}\biggr]+{1\over 2 e^2}{\sin^2 f\over
r^2}
\biggl [{\sin^2 f\over r^2} + 2\biggl({\partial f\over \partial
r}\biggr)^2\biggr] \biggr\} }$$
$$\Lambda = (ef_\pi )^3\int{r^2 dr \sin^2 f\biggl[ 1+ {4\over (ef_\pi)^2 }
\biggl(f'^2+ {\sin^2 f\over r^2}\biggr)\biggr] }$$
and finally the interesting term
$$T_{int} = 2 f_\pi^2 \kappa^2\, F_{ia}^{1}\, F_{jb}^{2}\, {4\pi\over d}
(\delta^{ij}-\hat d^i \hat d^j) \,D_{ab}(A_1^{\dagger} A_2)$$
where
$$F_{ia}^1 = - \beta^1 \dot \beta_a^1 + \dot \beta_i^1 \beta_a^1 -
\epsilon_{iap} \,(\dot \beta_b^1 \alpha^1 - \beta_b^1 \dot \alpha^1)$$
$$A_1 = \alpha^1 + i \,\vec \beta^1\cdot\vec \tau$$
$$(\alpha^1)^2 + |\vec\beta^1|^2 = 1$$
(correspondingly for $A_2$), and $\kappa$ is determined by
$$f(r) \sim {\kappa\over r^2}$$
and
$$\vec d = \vec R_1 - \vec R_2$$
$$d = |\vec d|.$$
The metric can be easily be obtained from this expression by choosing local
coordinates on the product ansatz manifold $(\vec R_1, \vec R_2, \vec\beta^1,
\vec\beta^2)$ and extracting the quadratic form relating their time
derivatives.

The potential$^7$ between two Skyrmions can be calculated to give
$$V= 4\pi f_\pi^2 \kappa^2{(1-\cos\theta)(3 \,(\hat n\cdot\hat d)^2 -1)
\over d^3}$$
where $\theta$, $\hat n$ pick out the element of $SU(2)$ given by $A_1
A_2^\dagger$.

The potential is clearly of higher order than the metric, hence the
dominant contribution to the scattering at large separation comes only from the
metric. Thus to leading order we may even neglect the potential, and then the
problem reduces to calculating the geodesics on the product ansatz manifold. We
are presently working this out.
\vskip 2.0cm
\centerline{\bf Aknowlegements}
\vskip 1.0cm
We thank M. Temple-Raston for useful discussions. This work supported in part
by NSERC of Canada and FCAR of Qu\'ebec.

\vfill\break

\vskip 1.0cm
\centerline{\bf References}
\vskip 2.0cm

\noindent{[1] N.S. Manton, {\it Phys. Rev. Lett.} {\bf 60} 1916 (1988)}
\vskip 0.3cm
\noindent{[2] G.W. Gibbons, N. S. Manton, {\it Nucl. Phys.} {\bf B274} 183
(1986)}
\vskip 0.3cm
\noindent{[3] T.M. Samols, {\it Comm. Math. Phys.} {\underbar{145}}, 149
(1992)}
\vskip 0.3cm
\noindent{[4] T.H.R. Skyrme, {\it Proc. Roy. Soc. Lon.}
{\underbar{260}}, 127 (1961)}
\vskip 0.3cm
\noindent{[5] E. Witten, {\it Nucl. Phys.} {\bf B223}, 422, 433
(1983)}
\vskip 0.3cm
\noindent{[6] M. Kutschera, C.J. Pethick, {\it Nucl. Phys.} {\bf A440} 670
(1985)}
\vskip 0.3cm
\noindent{[7] K. Isler, J. LeTourneux, M.B. Paranjape, {\it Phys. Rev.} {\bf
D43}, 1366 (1991), and references herein.}
\vskip 0.3cm
\noindent{[8] T.S. Walhout and J. Wambach, {\it Phys. Rev. Lett.}
{\underbar {67}}, 314 (1991)}
\vskip 0.3cm
\noindent{[9] B. Schroers, Durham preprint, DTP 93-29, hep-th/9308017, has
found similar behaviour for the case of rigidly spinning Skyrmions.}
\end